\documentclass[showpacs,superscriptaddress,showpacs,showkeys,nofootinbib,superscriptaddress,10pt]{revtex4}
\usepackage{amssymb,amsmath}
\usepackage{graphicx}
\usepackage{placeins}
\usepackage[usenames,dvipsnames]{color}
\usepackage{epstopdf}
\usepackage{subfigure}
\usepackage{parskip}
\usepackage[compact]{titlesec}
\usepackage[none]{hyphenat}
\usepackage[hyperfootnotes=false]{hyperref}
\usepackage[hyperfootnotes=false]{hyperref}
\usepackage{color}
\hypersetup{colorlinks=true,citecolor=blue,linkcolor=Blue,urlcolor=blue}

\DeclareMathAlphabet{\mathbfit}{OML}{cmm}{b}{it}
\newcommand{\ef}{\operatorname{Erf}}

\titlespacing{\section}{0pt}{1em}{0em}
\titlespacing{\subsection}{0pt}{1em}{0em}

\begin{document}
\title{\hfill {\footnotesize J. Phys. A 47, 285302 (2014)}\\ \vspace{1.25 cm}
{\large Dirac electron in graphene under supersymmetry generated magnetic fields} \vspace{.5 cm}}
\author{{\bf Bikashkali Midya}}
\email{bmidya@ulb.ac.be}
\affiliation{Physique Nucl\'eaire et Physique Quantique, Universit\'e Libre de Bruxelles, 1050 Brussels, Belgium.}
\author{{\bf David J Fern\'andez}}
\email{david@fis.cinvestav.mx}
\affiliation{Departamento de F\'isica, Cinvestav, A.P. 14 - 740, 07000 M\'exico D.F., Mexico.\vspace{.5 cm}}

\begin{abstract}
We use supersymmetry transformations to obtain new one parameter family of inhomogeneous magnetic fields $\mathbfit{B} = \widetilde{\mathcal{B}}(x,\lambda) \hat{e}_z$ for which the massless Dirac electron possesses exact solution. The inhomogeneity appearing in $\widetilde{\mathcal{B}}(x,\lambda)$ can be controlled by the parameter $\lambda$. The obtained magnetic fields are interpreted as deformed variants of some physically attainable well known magnetic fields. A particular example, when a constant magnetic field is deformed, is considered to show that equidistant Landau levels exist even in the presence of an infinite number of specially designed inhomogeneous magnetic fields. 

\end{abstract}
\pacs{72.80.Vp, 03.65.Pm, 11.30.Pb}
\keywords{Dirac-Weyl equation; Supersymmetric Quantum Mechanics; Dirac electron; Graphene }
 \maketitle

\section{Introduction}
In recent years, interest in studying graphene has increased steeply due to its exceptional
electronic properties and potentially significant applications in nanotechnology. Graphene, a single layer of carbon atoms arranged in a hexagonal honeycomb lattices, is the thinnest known
material in the universe and the strongest ever measured \cite{No+04,Ne+09,Ka12}. In the low energy approximation, the tight-binding description of the electron transport in graphene is reduced
to the massless Dirac-Weyl equation with Fermi velocity $v_F \sim c/300$. The relatively low Fermi velocity and the gapless linear energy dispersion near the Dirac points $K$ and $K'$, enabled one to simulate relativistic effects like anomalous Landau-Hall effect \cite{ZTSK05,WDM07,Bo+09}, the Zitterbewegung and the Klein paradox \cite{Ka06,KNG06,RZ08,SHG09}, in condensed matter systems. The optical-like
behavior of electron waves in graphene-based waveguide structure has also been
experimentally investigated \cite{HRP10,MIL11,ZHC09}. These phenomena are expected to play a needful role in the future fabrication of desirable nanoelectronic devices.

Since Dirac electrons cannot be confined by
electrostatic potentials due to Klein’s paradox, it was suggested
that magnetic confinement should be considered \cite{PCG06,MAE07,MI09,BFN10}. Inhomogeneous magnetic fields on the nanometer scale
can suppress the Klein tunneling and produce graphene based
structures to confine Dirac fermions, but the parameters
of the magnetic barriers cannot be changed fast enough. Electron confinement in the single \cite{MAE07}, double \cite{OR+08}, and multiple \cite{AM09,MVMP08}
square-well magnetic barriers and in delta
function magnetic barriers \cite{GS09} have been investigated. Heaviside-step-function-like magnetic
profile is used in \cite{HHYL12} to study the magnetic induced
waveguide structure. And it is shown that  the
guided modes are strongly confined by this gradually varied
magnetic field. 

Recently a series of studies concerning the
interaction of graphene electrons moving in magnetic
fields perpendicular to the graphene surface \cite{JM11,Gh09} and/or including electrostatic fields parallel to the surface \cite{PC07} have been carried out in order to find a way for
confining the charges. In all these works, the Dirac-Weyl
equation is considered, where a minimal coupling with the vector potential
describes the interaction with the external field. In general,
some kinds of numerical computation or semiclassical methods were needed to find the
energy levels of confined states or transmission coefficients for
scattering states under arbitrary magnetic fields. However, attempts have been made in \cite{SH08,KNN09} to investigate the same by analytical method like supersymmetric quantum mechanics (SUSY-QM) \cite{CKS95,JR98,mr04,Fe10,ai12}. This was possible due to the fact that the Hamiltonian of graphene in a background static magnetic field can be mathematically cast into one with quantum mechanical factorization properties used in SUSY-QM. It is shown that except for the ground state, the Hamiltonians describing the Dirac fermions in the two adjacent sites in the unit cell of the graphene lattice, have the same spectrum. Using the factorization technique, the exact expressions of discrete spectra and wave functions of graphene Dirac electron in various magnetic fields having translational symmetry are obtained. In all these cases the effective partner Hamiltonians are essentially shape invariant \cite{SI}. It is worth mentioning here that SUSY-QM has been applied as well to solvable models in the low energy approximation of carbon nanotubes \cite{LN03,Vit11,Vit14} and spherical molecules including fullerenes \cite{Vit13}. In ref. \cite{LN03}, the effect of homogeneous transverse magnetic fields on the energy spectrum of a carbon nanotube is explained. Whereas, the construction of new solvable systems within the SUSY framework is considered in \cite{Vit11} where the results are relevant not only for carbon nanotubes but also for graphene with potential barriers that possess translational invariance in one direction. A class of exactly solvable metallic and maximally semiconducting nanotubes in the presence of an external inhomogeneous magnetic field is shown to possess nontrivial integrals of motion, which establishes the $N=2$ nonlinear supersymmetry \cite{Vit14}. The nonlinear hidden supersymmetry generated by local supercharges is also shown to appear naturally in the finite-gap and time-reversal invariant configurations of twisted nanotubes \cite{CJ13}.

In this paper, we will obtain new one parameter families of inhomogeneous external magnetic fields $\widetilde{\mathcal{B}}(x,\lambda)$ for which the Dirac electron possesses exact solutions. The inhomogeneity appearing in $\widetilde{\mathcal{B}}(x,\lambda)$ can be controlled by the parameter $\lambda$. To this aim we revisit supersymmetric factorization approach to Dirac-Weyl equation and precisely exploit the nonuniqueness property of the superpotential (which is directly related to the vector potential, see section \ref{s2}). The one parameter family of magnetic fields so obtained are shown to be the deformed variants of some well known physically interesting magnetic fields. The corresponding effective Hamiltonians are no longer shape invariant and have altogether different functional forms with interesting features. The effect of these deformations on the energy spectrum and the probability current density associated with the lowest energy states of the Dirac electron shall also be investigated. For illustration purposes we will consider two examples: the Dirac electron being placed first in a harmonic well associated to a uniform magnetic field, and then in a Morse oscillator characterizing an exponentially decaying magnetic field. It is shown that the deformation of a uniform magnetic field enables one to generate the gradually varying magnetic fields which were achieved in \cite{HHYL12} earlier with the help of Heaviside-step function.

The paper is organized as follows. In section \ref{s2}, a brief overview of the quantum mechanical treatment of the Dirac-Weyl equation for Dirac fermion and its connection to first order SUSY-QM will be done. Next in section \ref{s3}, the most general form of the magnetic fields, bearing exactly solvable effective potentials, are going to be derived. Section \ref{s4} deals with two concrete examples: deformation of uniform and
exponentially decaying magnetic fields, respectively. Summary and future prospects concerning the possible experimental
implication of the work are stated in section \ref{s5}.

\section{SUSY-QM approach to Dirac-Weyl equation}\label{s2}
We consider Dirac electron moving in graphene under static
orbital magnetic field acting perpendicular to the graphene $x-y$ plane. Moreover, we disregard the electronic spin degree of freedom and the presence of any disorder in the system. Under these circumstances the low energy excitation of the electron around a Dirac point in the Brillouin zone is described by the following Dirac-Weyl equation \cite{MAE07}
\begin{equation}
\mathbfit{H} \Psi(x,y) = \upsilon_{F} \boldsymbol{\sigma}.\left[\mathbfit{p}+\frac{e \mathbfit{A}}{c}\right] \Psi(x,y) = E \Psi(x,y), \label{e2}
\end{equation}
where the Fermi velocity $v_F \sim 8\times 10^5 m/s$, $\boldsymbol{\sigma} = (\sigma_x,\sigma_y)$ are the Pauli matrices, and $\mathbfit{p} = -i \hbar(\partial_x,\partial_y)^T$ is the
two dimensional momentum operator. Here $(-e)$ is the charge of the electron and the vector potential $\mathbfit{A}$ is related to a slowly varying magnetic field by $\mathbfit{B}= \nabla \times \mathbfit{A}.$ There is a lot of freedom in choosing $\mathbfit{A}$ and a convenient choice is Landau gauge $\mathbfit{A} =  \mathcal{A}(x) \hat{e}_y, ~\mathbfit{B} =  \mathcal{B}(x) \hat{e}_z, ~\mathcal{B}(x) = \mathcal{A}'(x)$. Now, taking into account the translational invariance along (say) $y$ direction we assume the 
two component spinor $\Psi(x,y)$ in the form
\begin{equation}
\Psi(x,y) = e^{i k y} \left[
                        \begin{array}{c}
                          \psi^+(x) \\
                          i \psi^-(x) \\
                        \end{array}
                      \right],\label{e11}
\end{equation}
$k$ being the wave number in the $y$ direction and $\psi^\pm(x)$ describe the electron amplitude on two adjacent sites in the unit cell of graphene. Consequently, the Dirac-Weyl equation (\ref{e2}) yields two coupled first order linear differential equations
\begin{equation}
\left(\pm \frac{d}{dx} + \frac{e}{c \hbar} \mathcal{A} + k \right) \psi^\mp(x) = \frac{E}{\hbar \upsilon_F} \psi^\pm(x),\label{e3}
\end{equation}
which readily can be decoupled to get two non-relativistic Schr\"odinger equations $H^\pm \psi^\pm(x) = \mathcal{E} \psi^\pm(x)$, where 
\begin{equation}
H^\pm = -\frac{d^2}{dx^2} + V^\pm, ~~ V^\pm = \left(\frac{e \mathcal{A}}{c \hbar} + k\right)^2 \pm \frac{e}{c \hbar} \frac{d \mathcal{A}}{dx},\label{e4}
\end{equation}
\begin{equation}
\hspace{-4 cm} \mbox{and} ~~~~~~~~~~~~ \mathcal{E} = \frac{E^2}{\hbar^2 \upsilon_F^2}.\label{e10}
\end{equation}
Hence, the determination of a complete solution of the Dirac-Weyl equation is now reduced to the problem of solving the above two Schr\"odinger equations. Indeed, the very form of the Hamiltonians $H^\pm$ in equation (\ref{e4}) indicates that they can be embraced by the formalism of SUSY-QM if one defines the intertwining operators   
\begin{equation}
L^\pm = \mp \frac{d}{dx} + W(x), ~~~ W(x) = \frac{e \mathcal{A}(x)}{c \hbar} + k,\label{e40}
\end{equation}
where $W(x)$ is the superpotential. Now, for the two operators $L^\pm$, the equations in (\ref{e3}) reduce to
\begin{equation}
L^- \psi^-(x) = \sqrt{\mathcal{E}} \psi^+(x), ~~~~ L^+ \psi^+(x) =\sqrt{\mathcal{E}} \psi^-(x),\label{e15}
\end{equation}
respectively and the two Hamiltonians $H^\pm$ become factorized in the following way
\begin{equation}
H^\pm = L^\mp L^\pm, ~~~~~ V^\pm(x) = W^2 \pm W'.\label{e16}
\end{equation}
Equations (\ref{e15}) and (\ref{e16}) imply that $H^\pm$ are supersymmetric partners connected by the following intertwining relationships
$H^\pm L^\mp = L^\mp H^\mp$. This means that knowing all the eigenfunctions of $H^+$ one can determine the ones of $H^-$ using the operator $L^+$, and vice versa, using $L^-$ one can construct all the eigenfunctions of $H^+$ from those of $H^-$ except possibly for the ground state. Specifically, in standard unbroken first order supersymmetric quantum mechanics the eigenvalues and eigenfunctions of the two partner Hamiltonians are related by \cite{CKS95}:
\begin{eqnarray}
 \mathcal{E}_{n}^+ = \mathcal{E}_{n+1}^-, \qquad \mathcal{E}_{0}^- = 0, ~~~~ \psi_{0}^- \sim e^{-\int W(x) dx} ,\nonumber \\
 \psi_{n}^+ = \frac{L^- \psi_{n+1}^-}{\sqrt{\mathcal{E}_{n+1}^-}}, ~~~ \psi_{n+1}^- = \frac{L^+ \psi_{n}^+}{\sqrt{
\mathcal{E}_{n}^+}}, ~~~ n=0,1,2,\dots \label{e7}
\end{eqnarray}

At this point it should be mentioned that the superpotential $W(x)= - \psi_{0}^{-'}/\psi_{0}^-$ is related directly to the vector potential [see equation~(\ref{e40})] and hence to the magnetic field by the relation $\mathcal{B}(x) = \frac{c \hbar}{e} \frac{dW}{dx}$. Thus, for a particular choice of magnetic field the problem of determining the energy spectrum and the corresponding spinor wave functions of the Dirac electron reduces to solving the Schr\"odinger eigenvalue equation for either $H^+$ or $H^-$. And once we have the complete solutions of $H^+$ or $H^-$, we can obtain the solution of the Dirac electron by using equations (\ref{e11}), (\ref{e10}) and (\ref{e7}). So far, this technique has been implemented in order to
solve the Dirac-Weyl equation under several magnetic fields which give rise to the so-called shape invariant effective potentials $V^\pm(x)$. In the next section, departing from these cases we are going to generate new inhomogeneous magnetic fields for which the corresponding effective potentials are no longer shape invariant yet the Dirac-Weyl equation is exactly solvable.

\section{Dirac electron in SUSY deformed magnetic fields}\label{s3}
Here we will obtain the most general form of the magnetic field departing from a given initial one $\mathcal{B}(x)$, keeping intact the exact solvability of the Dirac-Weyl equation. To this end, we introduce a generalized superpotential\footnote{Henceforth we adopt the notations $F$ for the usual quantities and $\widetilde{F}$ for those which correspond to the deformation.}
\begin{equation}
\widetilde{W}(x) = W(x) + f(x)\label{e13}
 \end{equation}
which satisfies now a modified version of the Riccati equation (\ref{e16})
\begin{equation} \label{e12}
\widetilde W^2 + \widetilde W' = V^+(x) + \delta, \qquad \delta\in{\mathbb R},
\end{equation}
and whose solutions are going to generate new deformed magnetic fields
\begin{equation}
 \widetilde{\mathcal{B}} = \frac{c \hbar}{e} ~ \frac{d \widetilde{W}}{dx} =  \mathcal{B}(x) + \frac{c \hbar}{e} f'(x) \label{e14}
  \end{equation}
so that the corresponding Dirac-Weyl equation is once again exactly solvable. Besides, as it will be shown below, the effective potential $V^-(x)$ becomes now a one (or two) parameter family of potentials $\widetilde{V}^-$ which can be considered as deformations of $V^-$.  In general, for arbitrary $\delta$ the Riccati equation (\ref{e12}) is not always easy to solve to find the unknown function $f(x)$. In the following we first will consider $\delta = 0$; the case of $\delta \ne 0$ will be discussed subsequently. \\

{$\square$ {\bf {$\delta = 0$ (one parameter family of magnetic fields):}}} for $\delta = 0$ the equation (\ref{e12}) reduces to a Bernoulli differential equation whose solution is $f(x) = (\psi_0^-)^2 \left[\lambda +
\int_{-\infty}^x(\psi_{0}^-)^2 dt\right]^{-1}$ and hence
\begin{equation}
\widetilde{W}(x,\lambda) = - \frac{{\psi^-_0}'}{\psi_{0}^-} + \frac{(\psi_{0}^-)^2}{\lambda +
\int_{-\infty}^x(\psi_{0}^-)^2 dt},
\end{equation}
where $\lambda$ is an arbitrary real constant and $\psi_{0}^-(x)$ is the ground state eigenfunction of $V^-(x)$.  The partner potentials corresponding to the generalized superpotential $\widetilde{W}(x,\lambda)$ become now
\begin{equation}
\begin{array}{lll}
\widetilde{V}^+(x) = \widetilde{W}^2 + \widetilde{W}' = V^+(x), \\
\widetilde{V}^-(x,\lambda) = \widetilde{W}^2- \widetilde{W}'\\
\quad \quad \quad \quad  = V^-(x) - 2 \partial_{xx} \ln \left[\lambda + \int^x_{-\infty} (\psi_0^-)^2 dt\right],
\end{array}\label{e22}
\end{equation}
and the corresponding one parameter family of magnetic fields $\widetilde{\mathcal{B}}(x,\lambda)$ can be obtained using equation (\ref{e14})
\begin{equation}
\widetilde{\mathcal{B}}(x,\lambda) = \mathcal{B}(x) + \frac{c \hbar}{e} ~\partial_{xx}\ln \left[\lambda + \int_{-\infty}^{x} (\psi_0^-)^2 dt\right]. \label{e31}
\end{equation}

Several remarks are in order at this stage:
\begin{itemize}
\item[$\circ$] {The family of magnetic fields $\widetilde{\mathcal{B}}(x,\lambda)$ can be considered as a deformation of the usual magnetic field $\mathcal{B}(x)$ by an amount proportional to $f'(x)$. This infinite family of deformed magnetic fields generates an infinite family of effective partner potentials, $V^+(x)$ and $\widetilde{V}^-(x,\lambda)$, which in general are not longer shape invariant yet exactly solvable. The shape invariance property is retained only in the limit $\lambda \rightarrow \pm \infty$. In this limit both the effective potential and the magnetic field reduce to the undeformed ones, which shows that $V^-(x)$ and $\mathcal{B}(x)$ themselves are members of $\widetilde{\mathcal{B}}(x,\lambda)$ and $\widetilde{V}^-(x,\lambda)$, respectively.}

\item[$\circ$] {$\widetilde{\mathcal{B}}(x,\lambda)$ is in general inhomogeneous in nature, where the amount of additional inhomogeneity can be controlled by the parameter $\lambda$. Most interestingly, if the initially given magnetic field $\mathcal{B}(x)$ is uniform, the deformed magnetic fields $\widetilde{\mathcal{B}}(x,\lambda)$ become inhomogeneous. This case will be discussed in detail in section \ref{s4.1}. }

\item[$\circ$] {Although $\lambda$ is an arbitrary real number, one has to avoid those values of $\lambda$ for which both the magnetic field $\widetilde{\mathcal{B}}$ and the potential $\widetilde{V}^-$ become unphysical (singular). Since the values of the integral $\int_{-\infty}^x {\psi_0^-(t)}^2 dt$ lie between $0$ and $1$, the only acceptable values are $\lambda<-1$ or $\lambda>0$.}

\item[$\circ$] {All the potentials $\widetilde{V}^-(x,\lambda)$ have the same partner $V^+(x)$, so they are isospectral to $V^-(x)$, that is, $\tilde{\mathcal{E}}_n^- = \mathcal{E}_n^-$. The bound state wave functions $\widetilde{\psi}^-_n(x,\lambda)$ are obtained from
$\psi_n^+(x)$, and vice versa, using relations (\ref{e7}) with the modified intertwining operators $\tilde{L}^\pm = \mp \frac{d}{dx} + \widetilde{W}(x)$. Specifically,
\begin{equation}
\begin{array}{ll}
\widetilde{\psi}^-_0(x,\lambda) = \frac{\sqrt{\lambda(\lambda+1)}}{\lambda + \int_{-\infty}^x {\psi_0^-(t)}^2 dt} \psi_0^-(x),\\
\widetilde{\psi}_{n+1}^-(x,\lambda)= \psi_{n+1}^- + \frac{\psi_0^- ~ W[\psi_0^-,\psi_{n+1}^-]}{\mathcal{E}_{n+1}^-[\lambda+\int_{-\infty}^{x} \psi_0^-(t)^2 dt]},
\end{array}\label{e33}
\end{equation}
where $W[f_1 , f_2] = f_1 f_2'- f_2 f_1'$ is the Wronskian of the two functions $f_1(x), \ f_2(x)$. }  

\item[$\circ$] {For $\lambda=0$ and $-1$, the ground state wave function $\widetilde{\psi}_0^-$ is no longer normalizable. Therefore, $\tilde{\mathcal{E}}_0^-$ should be removed from the spectrum of $\widetilde{V}^-(x,\lambda)$. As a result, supersymmetry is broken in these two limit cases.}
\end{itemize}
Hence, for $\delta=0$ and $\lambda \in \mathbf{R}-\{[-1,0]\}$, the energy spectrum and spinor wave function of the Dirac electron moving in graphene under the supersymmetry deformed magnetic field $\widetilde{\mathcal{B}}(x,\lambda)$ are given by 
\begin{equation}
 \textstyle \begin{array}{ll}
  \widetilde{E}_0 = \hbar v_F~ \sqrt{{\mathcal{E}}_{0}^-} = 0, ~~~~ \widetilde{E}_{n+1} = \pm \hbar v_F~ \sqrt{{\mathcal{E}}_{n+1}^-}  \\
\widetilde{\Psi}_0 = e^{i k y}
  \left[\begin{array}{c}
                            0 \\
                            i ~ \widetilde{\psi}_{0}^-
                          \end{array}\right] , ~~ \widetilde{\Psi}_{n+1} = e^{i k y}
  \left[\begin{array}{c}
                            {\psi}_{n}^+ \\
                            i ~ \widetilde{\psi}_{n+1}^-
                          \end{array}\right].
 \end{array}\label{e34}
\end{equation}
Note that the energy spectrum is not altered by the deformation, and it contains both positive and negative eigenvalues which correspond to the electron and hole respectively, in a symmetric way. To probe the effect of the deformation on the nature of a massless Dirac electron, one should investigate the probability distribution $\tilde{\rho}_n(x)$ and current density $\tilde{j}_n(x)$. For a given state $\widetilde{\Psi}_n(x)$, they are determined by $\tilde{\rho}_n(x) = \widetilde{\Psi}_n^\dag \widetilde{\Psi}_n$ and $\tilde{j}_n = e v_F \widetilde{\Psi}_n^\dag \sigma_y \widetilde{\Psi}_n$ respectively. Since $\tilde{j}_0 = 0$, the zero energy state does not carry any current irrespective of the magnetic field. Both the probability and current densities, which corresponds to stationary states, are independent of
$y$, due to the translational symmetry in this direction.\\

$\square$ {\bf {$\delta \ne 0$ (two parameter family of magnetic fields):}} The case with $\delta \ne 0$ is more general and can be investigated in an analogous way as for $\delta =0$. But in that case one has to solve the general Riccati equation (\ref{e12}) to evaluate $f(x)$. This nonlinear equation can be linearized by assuming $f(x) = u'(x)/u(x)$, where $u(x)$
satisfies the following homogeneous second order linear differential equation
\begin{equation}
u''(x) + 2 W(x) u'(x) - \delta u(x)= 0.\label{e17}
\end{equation}
The general solution of equation (\ref{e17}) is given by a linear combination of two linearly independent
solutions $u_{1,2}$ as $u(x) = \alpha u_1(x) + \beta u_2 (x)$. For all shape-invariant potentials, the equation (\ref{e17}) can
be reduced to a hypergeometric or confluent hypergeometric differential equation.
That is, the two solutions $u_{1,2}$ can be expressed in terms
of hypergeometric or confluent hypergeometric functions. Once the analytical form of $u(x)$ is found, the partner potentials $\widetilde{V}^\pm(x)$
corresponding to the generalized superpotential, $\widetilde{W}(x)$ ($= W + u'/u$) become now
\begin{equation}
\widetilde{V}^+(x) = V^+(x) + \delta, ~~~\widetilde{V}^-(x) = V^-(x) - 2 \partial_{xx} \ln u(x) +\delta,\label{e22}
\end{equation}
and the corresponding deformed magnetic field can be obtained using equation (\ref{e14}). Both the effective potential $\widetilde{V}^-$ and the magnetic field $\widetilde{\mathcal{B}}$ depend now on the parameter $\delta$ and the quotient $\alpha/\beta$ or $\beta/\alpha$. It is worth mentioning here that in order that the generalized superpotential, the deformed magnetic field and the effective potential 
$\widetilde{V}^-(x)$ are free from singularities, it is necessary that the function $u(x)$ should be nodeless. Hence, we have to choose 
$\delta$ and the parameters appearing in $u(x)$ such that $u(x)$ becomes either strictly positive or
strictly negative functions, which first of all implies that $\delta\geq -\mathcal{E}_{0}^+$. In addition,
once a $\delta$ consistent with this restriction has been chosen, we will have to analyze further the asymptotic
behavior of the function $u(x)$ in order to select the nodeless one.

The eigenfunctions and bound state energies of
$\widetilde{V}^+$ are easily obtained from those of $V^+(x)$, namely, $\widetilde\psi_{n}^+(x) =\psi_{n}^+(x), ~ \tilde{\mathcal{E}}_{n}^+ = {\mathcal{E}}_{n}^+ + \delta.$ On the other hand, the energies of  $\widetilde{V}^-$ are given by 
$\tilde{\mathcal{E}}_{n+1}^- = \tilde{\mathcal{E}}_{n}^+= \mathcal{E}_{n+1}^- +\delta$ and the corresponding eigenfunctions $\widetilde{\psi}^-_{n+1}$ are obtained from $\widetilde\psi_{n}^+(x)$ using relation (\ref{e7}) with the modified intertwining operator $\tilde{L}^+ = - \frac{d}{dx} + W(x) + \frac{u'(x)}{u(x)}$ replacing the unmodified one. Moreover, consistent with the fact that the state $\widetilde{\psi}_{0}^-(x)$ which is annihilated by $\tilde{L}^-$ could be either of square integrable or not, then the zero eigenvalue $\tilde{\mathcal{E}}_{0}^- = 0$ could be or not in the spectrum of $\widetilde H^-$ respectively \cite{Fe10}. Note that $\tilde{L}^- \widetilde{\psi}_{0}^-(x) = 0$ is a first-order differential equation with solution given by $\widetilde{\psi}_{0}^-(x) \sim \psi_{0}^-(x)/u(x).$ In contrast to the case with $\delta =0$, here the energy spectrum of the Dirac electron in the usual and deformed magnetic fields are not identical. Nonetheless, the square of the energy eigenvalues are related by $\widetilde{E}_n^2 = E_n^2 + \hbar^2 v_F^2 \delta.$ 

\section{Application}\label{s4}
It is worth mentioning that the mechanism discussed in the previous section is very general and can be applied to any magnetic field which initially yield exactly solvable shape-invariant effective potentials. However, for illustration purposes here we will consider just two examples: the Dirac electron being placed first in a harmonic well which feels a uniform magnetic field, and then in a Morse oscillator experiencing an exponentially decaying magnetic field. Since the analytical expressions of the deformed magnetic field, the corresponding effective potentials and the excited state wave functions are too involved for $\delta \ne 0$, we shall only consider the case with $\delta = 0$.

\subsection{Deformation of a uniform magnetic field}\label{s4.1}
Let us first choose a uniform magnetic field perpendicular to the graphene plane in the positive $z$ direction, $\mathbfit{\mathcal{B}} = \hat{e}_z B_0$, $B_0>0$. It can be generated by choosing the standard Landau gauge $\mathbfit{A} = \hat{e}_y B_0 x$. In this case the superpotential $W(x) =  \frac{\omega}{2} x + k$ yields the shape-invariant effective potentials $V^\pm (x) = \frac{\omega^2}{4} \left(x+\frac{2 k}{\omega}\right)^2 \pm \frac{\omega}{2}$, respectively. Here $\omega = \frac{2 e }{c \hbar} B_0$ whose dimension is (length)$^{-2}$. The eigenvalues and eigenfunctions of these shifted harmonic oscillators turn out to be
\begin{equation}\begin{array}{ll}
\mathcal{E}_{0}^- = 0, ~~~ \mathcal{E}_{n+1}^- = \mathcal{E}_{n}^+ = \omega(n+1), ~~~n= 0,1,2,\dots\\
\psi_{n}^- = \psi_{n}^+ = N_n~ e^{-\frac{\omega}{4}(x + \frac{2k}{\omega})^2} ~~ H_n\left[\sqrt{\frac{\omega}{2}}\left(x + \frac{2k}{\omega}\right)\right],
\end{array}\label{e18}
\end{equation}
where $N_n = \sqrt{\frac{1}{2^{n} n! } \left(\frac{\omega}{2\pi}\right)^{\frac{1}{2}}}$ and $H_n$ denotes the Hermite polynomial of order $n \in \mathbb{N}$. Hence, the complete solution of the corresponding Dirac-Weyl equation for the constant magnetic field can be found using equations (\ref{e11}), (\ref{e7}) and (\ref{e18}). The above example is the well known case of equidistant Landau levels \cite{LL65}.
\begin{figure*}[htb]
\includegraphics[width=7.5 cm,height=5 cm]{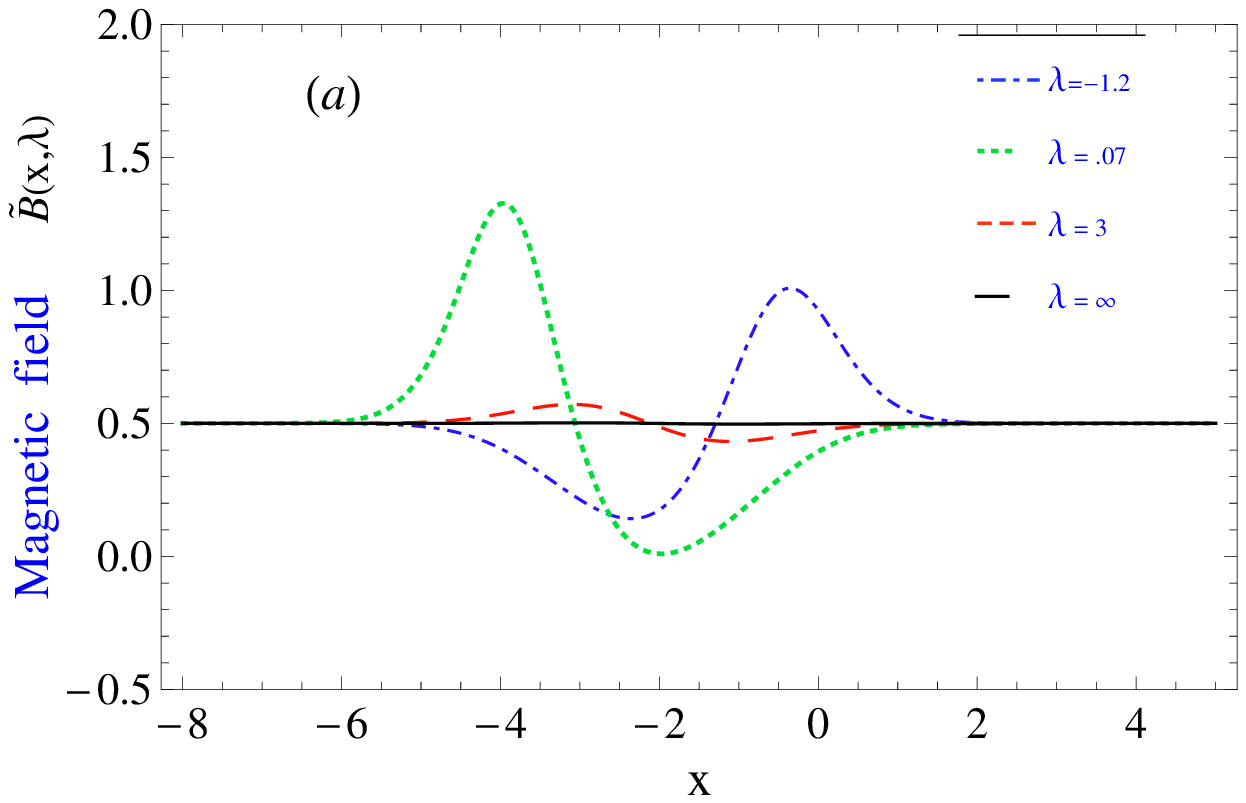}~ \includegraphics[width=7 cm, height=5 cm]{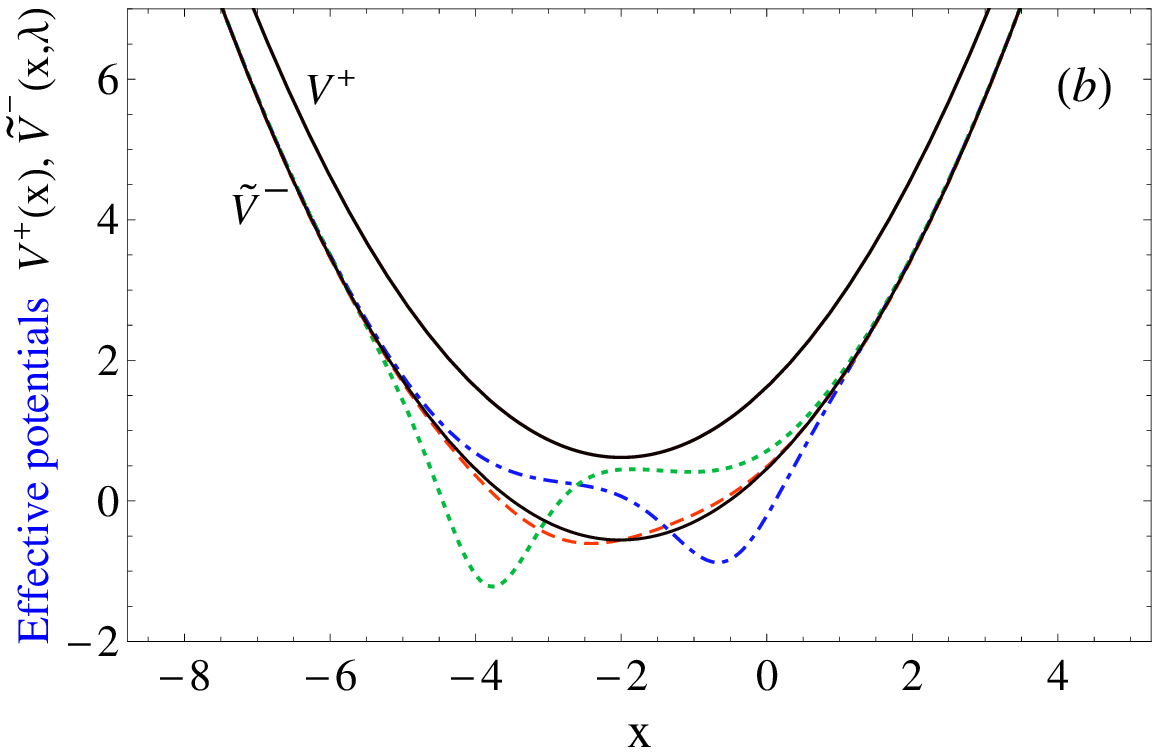}
\includegraphics[width=7.5 cm, height=4.5 cm]{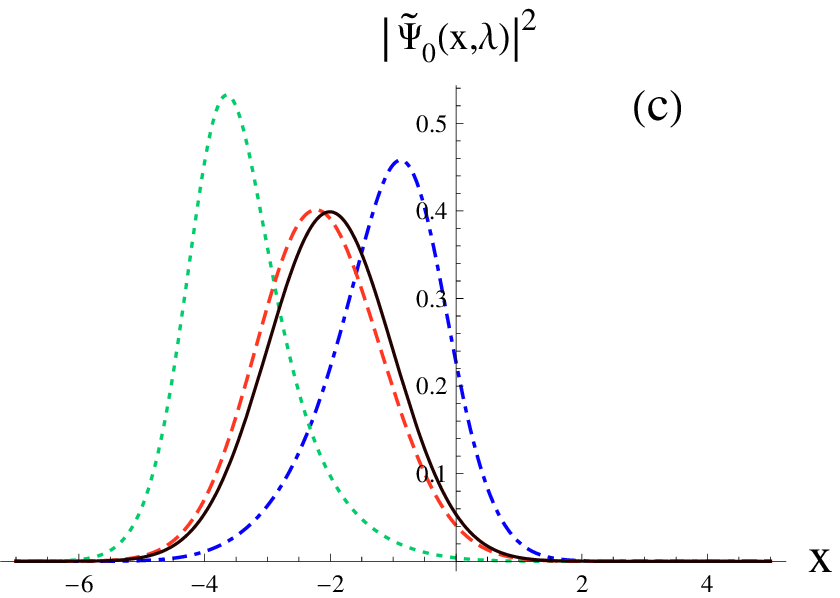} ~ \includegraphics[width=7 cm, height=4.5 cm]{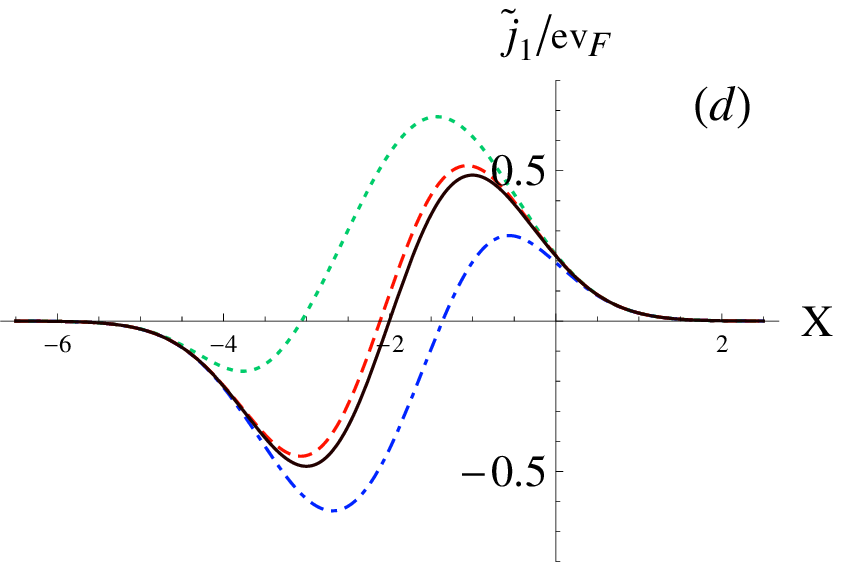}
\caption{(Color online) (a) Deformed magnetic fields $\widetilde{\mathcal{B}}(x,\lambda)$ for $\lambda = -1.2, 0.07, 3$ and $\infty$; (b) Effective potentials $\widetilde{V}^-(x,\lambda)$ and $V^+(x)$; (c) ground state probability densities $|\widetilde{\Psi}_0(x,\lambda)|^2$ for the same values of $\lambda$ as in (a); (d) corresponding current densities for the first excited state. Here we have considered $B_0 =.5, w=1$.}\label{f1}
\end{figure*}
\FloatBarrier

Now for $\delta =0$, the deformation of the uniform magnetic field can be generated using the normalized ground state wave function $\psi_0^-(x)$ and using equation (\ref{e31}). The explicit expression of the one parameter family of magnetic fields and effective potentials \cite{mi84} are obtained as
\begin{equation}
\textstyle \begin{array}{ll}
 \widetilde{B}(x,\lambda) = B_0 + \frac{2 B_0}{\omega} \partial_{xx} \ln\left( 2 \lambda + 1 + \ef\left[\sqrt{\frac{\omega}{2}}\left(x+\frac{2 k}{\omega}\right)\right]\right), \\
 
\widetilde{V}^-(x,\lambda) = V^- - 2 \partial_{xx} \ln\left( 2 \lambda + 1 + \ef\left[\sqrt{\frac{\omega}{2}}\left(x+\frac{2 k}{\omega}\right)\right]\right),
\end{array}\label{e32}
\end{equation}
respectively. In order to avoid the singularity in both the magnetic field and the potential one has to choose either $\lambda <-1$ or $\lambda>0$. Since the energy spectra of the potentials $\widetilde{V}^-$ remain the same as their undeformed version $V^-$, therefore the Landau levels still exist although the magnetic fields are no longer uniform.

The bound state wave functions of the potential $\widetilde{V}^-(x,\lambda)$ are obtained with the help of equation (\ref{e33})
\begin{equation}
\begin{array}{ll}
\widetilde{\psi}_0^- =  \frac{2 N_0 \sqrt{\lambda(\lambda+1)}  ~e^{-\frac{\omega}{4}(x + \frac{2k}{\omega})^2}}{2 \lambda + 1 + \ef\left[\sqrt{\frac{\omega}{2}}\left(x+\frac{2 k}{\omega}\right)\right]},\\
\widetilde{\psi}_{n+1}^- = N_{n+1} e^{-\frac{\omega}{4}(x + \frac{2k}{\omega})^2} \left[ H_{n+1}\left[\sqrt{\frac{\omega}{2}}\left(x + \frac{2k}{\omega}\right)\right] + \frac{2 e^{-\frac{\omega}{2}(x + \frac{2k}{\omega})^2}}{\sqrt{\pi}\left(2 \lambda + 1 + \ef\left[\sqrt{\frac{\omega}{2}}\left(x+\frac{2 k}{\omega}\right)\right]\right)}  H_n\left[\sqrt{\frac{\omega}{2}}\left(x + \frac{2k}{\omega}\right)\right]\right].
\end{array} \label{e44}
\end{equation}
Hence the complete solution, $\widetilde{E}_n$ and $\widetilde{\Psi}_n$, of the Dirac electron moving under the deformed magnetic field (\ref{e32}) can be obtained using equation (\ref{e34}). Interestingly, the energy spectrum $\widetilde{E}_{n+1} = \pm~ \hbar \upsilon_F  \sqrt{\omega (n+1)}$ together with $\widetilde{E}_0 = 0$, does not depend on the wave number $k$. 

In figures \ref{f1}(a) we have plotted the deformed magnetic fields for $\lambda =-1.2, 0.07, 3$ and $\infty$. The corresponding effective potentials $V^+(x)$ and $\widetilde{V}^-(x,\lambda)$ and ground state probability density are drawn in figures \ref{f1}(b) and \ref{f1}(c) respectively, while the current density $\tilde{j}_1/ev_F$ is shown in figure \ref{f1}(d). From these figures it is clear that the larger the value of $|\lambda|$ the smaller the effect of deformation. On the other hand, for a particular $\lambda$ the deformation is strong for small $x$, whereas the deformation is negligible for sufficiently large values of $x$. The position and value of the potential minima in $\widetilde{V}^-$ are also different relative to those of $V^-$. This has a direct influence on the position and value of the maxima of the ground state probability distribution. The changes of phase in the current density $\tilde{j}_1/ev_F$ are also due to the deformed magnetic fields. From figure \ref{f1}(a), it is also clear that the deformed magnetic fields are very similar to the gradually varying magnetic fields which were achieved earlier (see figure 1(b) of \cite{HHYL12}) with the help of Heaviside-step like function.

\subsection{Deformation of an exponentially decaying magnetic field}
In this case we consider a magnetic field which is exponentially decaying in the positive $x$-direction \cite{Gh09},
\begin{equation}
\mathcal{B}(x)  = B_0 e^{-x}, \quad B_0 >0.
\end{equation}
Hence, the non-zero component of the vector potential is given by $\mathcal{A}(x) = -B_0 e^{-x}$, which yields the superpotential $W(x) = k - D e^{-x}$ and shape-invariant Morse potential $V^\pm (x) = D^2 e^{-2 x} - D(2k \mp 1) e^{-x} + k^2$ with $D = \frac{e B_0}{\hbar c}$. The bound state solutions of these two potentials are given by \cite{CKS95}  
\begin{equation}
\begin{array}{ll}
\mathcal{E}_0^- = 0; ~ \mathcal{E}_{n+1}^- = \mathcal{E}_n^+ = (n+1)(2k-n-1), \\
\psi^\pm_{n} (x) = N_n^\pm ~ e^{-D e^{-x} - ({s}^\pm-n) x}  ~ L_n^{2s^\pm-2n}(2 D e^{-x}),
\end{array}
\end{equation}
where $N_n^\pm = (2D)^{s^\pm - n}  \sqrt{2(s^\pm-n)n!/(2 s^\pm - n)!}$, $s^- = k, s^+ = k - 1$, $n=0,1,2... \le k$, and $L_n^\alpha(x)$ are the associated Laguerre polynomials.

\begin{figure*}[htb]
\includegraphics[width=7.5 cm,height=5 cm]{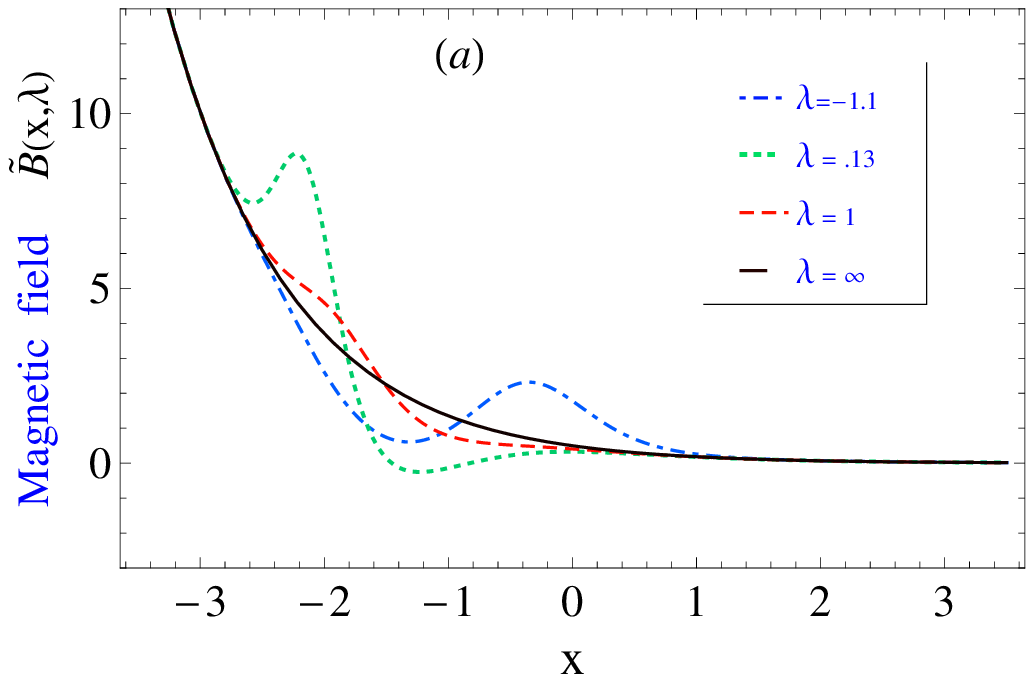}~ \includegraphics[width=7 cm, height=5 cm]{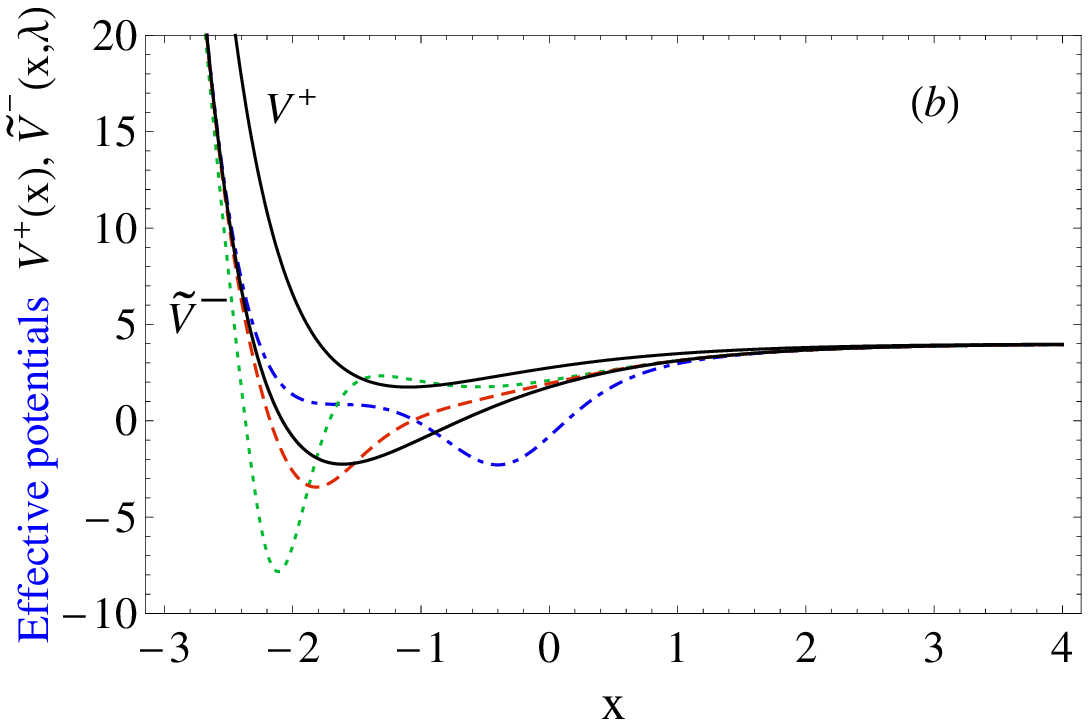}
\includegraphics[width=7.5 cm, height=4.5 cm]{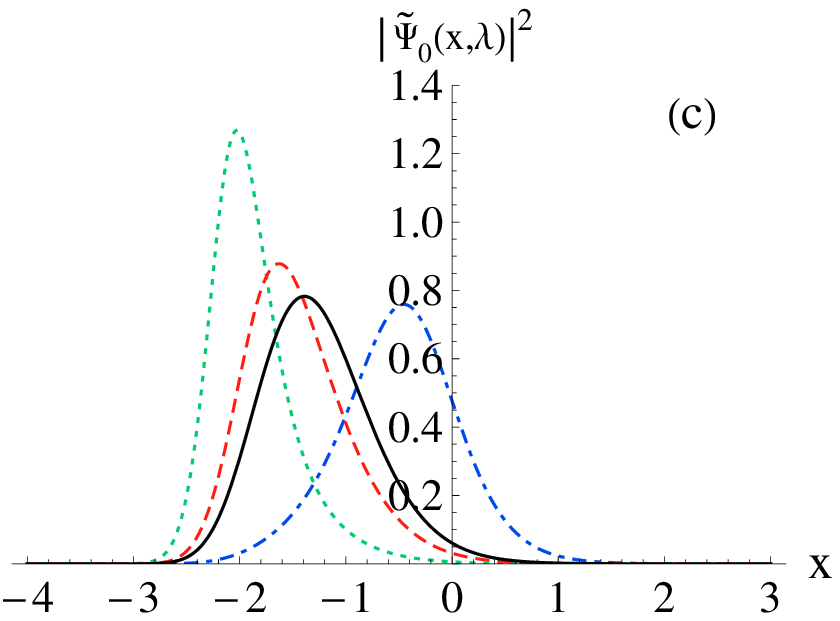} ~ \includegraphics[width=7 cm, height=4.5 cm]{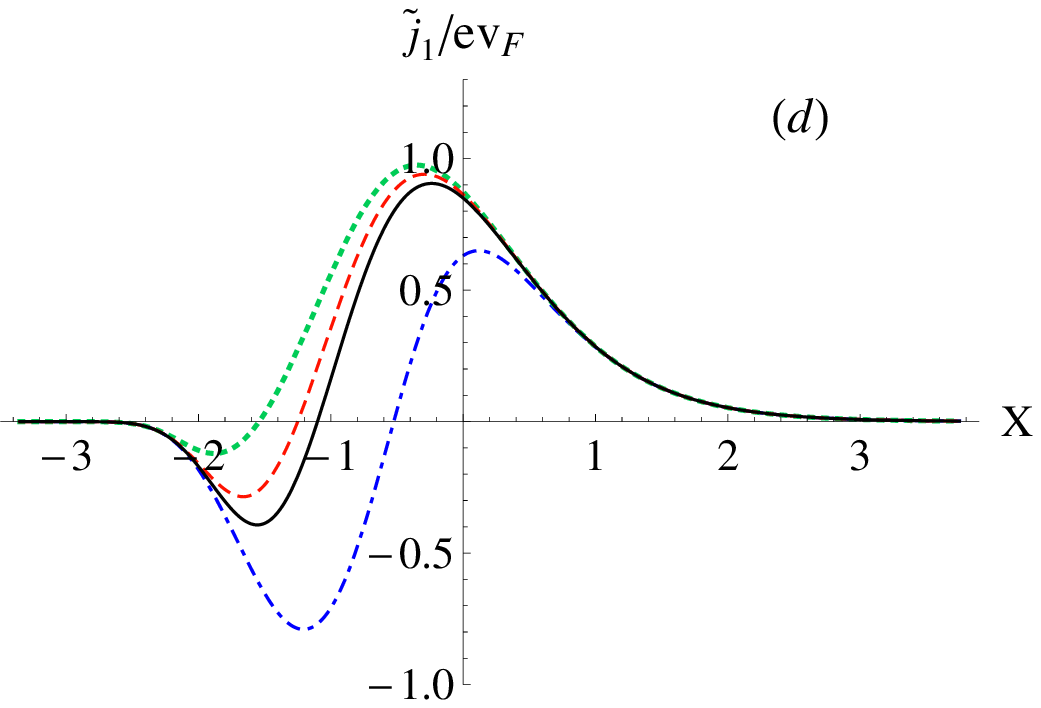}
\caption{(Color online) (a) Deformed magnetic fields $\widetilde{B}(x,\lambda)$ for $\lambda = -1.1, 0.13,  1,$ and $\infty$; (b) Effective potentials $\widetilde{V}^-(x,\lambda)$ and $V^+(x)$; (c) The ground state probability densities for the same values of $\lambda$ as in (a); (d) Corresponding current densities for the first excited state. Here we have considered $D=.5, k=2.$}\label{f2}
\end{figure*}
\FloatBarrier

Here the form of the one parameter family of deformed exponentially decaying magnetic fields and corresponding effective potentials is given by
\begin{equation*}
\begin{array}{ll}
\widetilde{\mathcal{B}}(x,\lambda) = B_0 e^{-x} + \frac{B_0}{D} ~\partial_{xx} \ln \left[ \lambda  (2k-1)! + \Gamma(2k,2D e^{-x})\right],\\
\widetilde{V}^-(x,\lambda) = V^-(x) -2\partial_{xx} \ln \left[ \lambda  (2k-1)! + \Gamma(2k,2D e^{-x})\right],
\end{array}
\end{equation*}
where $\Gamma(a,x) = \int_x^\infty e^{-t} t^{a-1}  dt$ is the incomplete Gamma function. The normalized bound state wave functions of $\widetilde{V}^-$ are given by :
\begin{equation*}
\begin{array}{ll}
\widetilde{\psi}_0^-(x,\lambda) =\frac{(2D)^k\sqrt{\lambda(\lambda+1)(2k-1)!} ~ e^{-D e^{-x} - k x}}{\lambda (2k-1)! + \Gamma(2k,2D e^{-x})}\\
\widetilde{\psi}_{n+1}^- = N_{n+1}^- e^{-D e^{-x}-(k-n)x}\left[e^x L_{n+1}^{2k-2n-2}(2D e^{-x})+ \frac{e^{-2D e^{-x}-2k x} \left[2D L_n^{2k-2n-1}(2D e^{-x})+ (n+1)e^x L_{n+1}^{2k-2n-2}(2D e^{-x})\right]}{(n+1)(2k-n-1)[\lambda (2k-1)! + \Gamma(2k,2D e^{-x})]}\right].
\end{array}
\end{equation*}
In figure \ref{f2}(a) we plot the deformed exponentially decaying magnetic fields for $\lambda =-1.1, 0.13, 1,$ and $\infty$. Whereas the effect of these deformed magnetic fields on the nature of effective potentials, ground state probability density and on the first excited state current density are shown in figure \ref{f2}(b),(c) and (d) respectively. 

\section{Summary}\label{s5}
To conclude, we have suggested a mechanism to find the most general form of a one parameter family of inhomogeneous magnetic field $\mathcal{\widetilde{B}}(x,\lambda)$ generated by applying the supersymmetric transformation to a given initial magnetic field $\mathcal{B}(x)$. It is shown that $\mathcal{\widetilde{B}}(x,\lambda)$ is a deformed variant of $\mathcal{B}(x)$. The discrete spectrum and eigenfunctions of the corresponding Dirac-Weyl equation are obtained analytically in closed form. The mechanism is illustrated with the help of two examples: the Dirac electron being placed first in a harmonic well associated to a uniform magnetic field, and then in a Morse oscillator characterizing an exponentially decaying magnetic field. The effects of the deformation on the probability and current densities are also investigated. Moreover, the construction of two parameter family of magnetic field is proposed. Since the aim of this work is to determine the most general form of the inhomogeneous magnetic fields for which the Dirac-Weyl equation can be solved exactly, a deeper investigation to uncover the new physics has been left for
further studies. However, as the inhomogeneity appearing in $\mathcal{\widetilde{B}}(x,\lambda)$ is controllable by the parameter $\lambda$, then the obtained magnetic fields can be used in graphene waveguide experiment of charge confinement. Note that the technology for doing this already exists as the profile of magnetic field can be generated experimentally by using ferromagnetic
gates located beneath the substrate on which the graphene layer is deposited \cite{WLLM11}. Hence we expect that apart from their
intrinsic mathematical interest, the results reported here will be useful for analyzing the Dirac electron confinement by inhomogeneous magnetic fields.

\section*{Acknowledgement}
\noindent BM thanks the Physics department of Cinvestav, Mexico, for a visit and hospitality at the outset of this work. He also acknowledges the postdoctoral grant from the Belgian Federal Science Policy Office co-funded by the Marie Curie Actions from the European Commission. DF acknowledges the financial support of Conacyt (M\'exico), project 152574.


\begin{thebibliography}{100}
\bibitem{SI} A Hamiltonian is said to be shape invariant, if its SUSY partner has the same spatial dependence as the original Hamiltonian with possibly altered
parameters. See ref. \cite{CKS95}.
\end{thebibliography}
\end{document}